\documentclass[aps,pre,twocolumn,groupedaddress]{revtex4}

\newif\ifpdf
\ifx\pdfoutput\undefined
\pdffalse 
\else
\pdfoutput=1 
\pdftrue
\fi

\ifpdf
\usepackage[pdftex]{graphicx}
\else
\usepackage{graphicx}
\fi
\usepackage{hyperref}

\begin{document}

\ifpdf
\DeclareGraphicsExtensions{.pdf, .jpg}
\else
\DeclareGraphicsExtensions{.eps, .jpg}
\fi

\def\hslash{\hbar}
\def\imag{i}
\def\grad{\vec{\nabla}}
\def\div{\vec{\nabla}\cdot}
\def\curl{\vec{\nabla}\times}
\def\DDt{\frac{d}{dt}}
\def\ddt{\frac{\partial}{\partial t}}
\def\ddx{\frac{\partial}{\partial x}}
\def\ddy{\frac{\partial}{\partial y}}
\def\lap{\nabla^{2}}
\def\divv{\vec{\nabla}\cdot\vec{v}}
\def\gradS{\vec{\nabla}S}
\def\vvec{\vec{v}}
\def\wc{\omega_{c}}
\def\<{\langle}
\def\>{\rangle}
\def\Tr{{\rm Tr}}
\def\Csch{{\rm csch}}
\def\Coth{{\rm coth}}
\def\Tanh{{\rm tanh}}
\def\g2{g^{(2)}}
\newcommand{\al}{\alpha}

\newcommand{\la}{\lambda}
\newcommand{\del}{\delta}
\newcommand{\om}{\omega}
\newcommand{\ep}{\epsilon}
\newcommand{\pd}{\partial}
\newcommand{\bra}{\langle}
\newcommand{\ket}{\rangle}
\newcommand{\bbra}{\langle \langle}
\newcommand{\kket}{\rangle \rangle}
\newcommand{\non}{\nonumber}
\newcommand{\be}{\begin{equation}}
\newcommand{\ee}{\end{equation}}
\newcommand{\bea}{\begin{eqnarray}}
\newcommand{\eea}{\end{eqnarray}}

\title{Exactly solvable approximating models for Rabi Hamiltonian dynamics}

\author{Andrey Pereverzev}
\email{aperever@mail.uh.edu}
\affiliation{Department of Chemistry and Center for Materials Chemistry, 
University of Houston \\ Houston, TX 77204}
\author{Eric R. Bittner}
\affiliation{Department of Chemistry and Center for Materials Chemistry, 
University of Houston \\ Houston, TX 77204}
\date{\today}

\begin{abstract}
The interaction between an atom and a one mode external driving field is 
an ubiquitous problem in many branches of physics and is 
often modeled using the Rabi Hamiltonian.  
In this paper we present a series of analytically solvable Hamiltonians that approximate 
 the Rabi Hamiltonian and compare our results to the Jaynes-Cummings model 
which neglects the so-called counter-rotating term in the Rabi Hamiltonian.  
Through a unitary transformation that diagonlizes the Jaynes-Cummings model, 
we transform the counter-rotating term 
into separate terms representing several different physical processes.  By keeping only certain 
terms, we can achieve an excellent approximation to the 
exact dynamics within specified parameter ranges.   

 \end{abstract}

\pacs{}

\maketitle

\section{Introduction}

The Rabi Hamiltonian is an elegant model for describing the transitions between 
two electronic states coupled linearly to a single mode of a harmonic driving field 
within the dipole approximation.  
Because of its simplicity in form, it plays an important role in many areas of 
physics from condensed matter physics and biophysics to quantum optics \cite{Allen,Leggett}. 
Given the apparent simplicity of this model and its wide range of applicability, it is not 
surprising that various aspects have been studied both analytically and numerically \cite{Wagner,Klimov,Yabuzaki,Kus2,Muller,Bonci1,Bonci2,Bishop,Feranchuk}
Remarkably, exact solutions have not been thus far presented except for special cases \cite{Kus1}
even though it has been suggested that the problem may be solved exactly \cite{Reik1,Reik2}
The Jaynes-Cummings model is a solvable approximation to the spin-boson model that neglects the 
counter-rotating term in the Rabi Hamiltonian \cite{Jaynes,Shore}. In general, it provides a reasonable
approximation to the course-grained dynamics in the limit of weak coupling and weak field. 
For stronger fields and couplings, however, the model breaks down.   While perturbative
treatments can be used to some extent \cite{Zubairy, Phoenix}, they give rise to fast oscillations 
and a dependence 
upon the phase of the initial state.   Furthermore, it seems rather dangerous to introduce a 
term as a perturbation which may be as strong as terms already present in the unperturbed 
Hamiltonian.  

In this paper, we present an alternative approach for including the counter-rotating terms into
the Jaynes-Cummings model.  We do this by transforming the counter-rotating term in
the Rabi Hamiltonian to the basis in which the Jaynes-Cummings term is diagonal and then
truncating the transformed counter-rotating operators to obtain a new series of exactly solvable models
that are related by various symmetry operations.  We then compare dynamics of the
excited state survival probability for our approximating models
to the Jaynes-Cummings model and to numerically exact solutions of the Rabi Hamiltonian 
and show that our approximate models do far better job in capturing both the long time 
decay and fine-structure in both the weak and strong field limits. 

The rest of this paper is organized as follows. In Sec.\ref{Hamiltonians} we obtain and
justify the approximating Hamiltonians. In Sec. \ref{Solution} eigenstates and eigenvalues
of these Hamiltonians are found. Excited state survival probability for one of the
approximating  Hamiltonians and its comparison to the results for the Jaynes-Cummings 
and Rabi Hamiltonains 
are given in Sec. \ref{Survival}.
 
\section{Obtaining approximating Hamiltonians}   \label{Hamiltonians}

The Rabi Hamiltonian describing interaction of a two level atom with a single-mode harmonic 
field can be written as ($\hbar=1$)
\be
H= \frac{\omega}{2}\sigma_z + \nu a^{\dagger}a + 
g(\sigma^++\sigma^-)(a+a^{\dagger}).  \label{Hamrabi}
\ee
Here $\sigma^+$ and $\sigma^-$ are spin-flip operators 
that satisfy 
\be
\sigma^-\sigma^++\sigma^+\sigma^-=1, \qquad \sigma^+\sigma^+=\sigma^-\sigma^-=0,
\ee
$\sigma_z=2\sigma^+\sigma^--1$, $a^{\dagger}$ and $a$ are the boson
creation and annihilation operators, and $a^{\dagger}a=\hat{n}$ is the boson number operator. 
Hamiltonian (\ref{Hamrabi}) can be split into two parts
as 
\be
H=H_{JC}+V,
\ee
where $H_{JC}$ is Jaynes-Cummings Hamiltonian 
\be
H_{JC}=\frac{\omega}{2}\sigma_z + \nu a^{\dagger}a + 
g(\sigma^+a+\sigma^-a^{\dagger}) \label{HJC}
\ee
and $V$ is the so-called counter-rotating term,
\be
V=g(\sigma^+a^{\dagger}+\sigma^-a).\label{potential}
\ee
$H_{JC}$ can be brought to a diagonal 
form, $\tilde{H}_{JC}$, by a suitable unitary 
transformation, 
\be
\tilde{H}_{JC}=UH_{JC}U^{-1},
\ee
in which $U$ is a unitary operator of the form \cite{Yu}
\be
U=e^{\sigma^+A-\sigma^-A^{\dagger}}. \label{u1}
\ee
In this paper we restrict our attention to the resonant case in which  $\nu=\om$ in Eq.(\ref{Hamrabi}).
For this case, $A$ has the following simple form
\be
A=\frac{\pi}{4\sqrt{\hat n+1}} a.
\ee
Here we use $\hat n$ for $a^\dagger a$ to simplify the notation.
The unitary transformation operator in Eq. (\ref{u1}) can be brought into the 
following useful form,
\be
U=\frac{1}{\sqrt{2}} + K(\hat n)(1-\sigma_z)
+\sigma^+ L(\hat n) a - \sigma^- a^{\dagger}L(\hat n) \label{unitary},
\ee
in which $K(\hat n)$ and $L(\hat n)$ are given by
\be
K(\hat n)=\left(\frac{\sqrt{2}-1}{2\sqrt{2}}\right)\delta(\hat n),
\qquad L(\hat n)=\frac{1}{\sqrt{2\hat n+2}}.\label{kl}
\ee
Here  $\delta(\hat n)$ is a projection operator on the ground state of the field.
Unitary transformation generated by $U$  diagonalizes  $H_{JC}$ as follows
\bea
\tilde{H}_{JC}&=&\frac{\omega}{2}\sigma_z + \omega {\hat n} \non \\
& &+\frac{g}{2}\left((\sigma_z+1)\sqrt{\hat n+1}
+(\sigma_z-1)\sqrt{\hat n}\right)
\eea
in which the eigenstates are given by
\be
|\theta_n^{\downarrow}\ket=|\downarrow\ket|n\ket, 
\qquad |\theta_n^{\uparrow}\ket=|\uparrow\ket|n\ket \label{jcstates}
\ee
where $|\downarrow\ket$ and $|\uparrow\ket$ are eigenstates of $\sigma_z$ 
with eigenvalues of $-1$ and $+1$, while $|n\ket$ is an eigenstate of $\hat n$
with eigenvalue $n$.

We now consider how the unitary transformation that diagonalizes $H_{JC}$ transforms
the total Hamiltonian (\ref{Hamrabi}). 
\be
\tilde{H}=UHU^{-1}=UH_{JC}U^{-1}+UVU^{-1} \label{Hamrabitilde}
\ee
The first term is diagonal and we focus our attention 
onto 
 $\tilde V=UVU^{-1}$.  $\tilde V$ can be written as a sum of four terms:
\be
UVU^{-1}=\tilde V=\tilde V_1+\tilde V_2 + \tilde V_3+\tilde V_4 \label{tildev}
\ee
where
\bea
\tilde V_1&=&g\left(\sigma^+F_1(\hat n)a^3+\sigma^-(a^{\dagger})^3F_1(\hat n)\right)\non \\
\tilde V_2&=&g\left(\sigma^+ a^{\dagger} F_2(\hat n)+\sigma^-F_2(\hat n)a \right)\non\\
\tilde V_3&=&g\left(\sigma_z\left( F_3(\hat n)a^2+(a^{\dagger})^2F_3(\hat n)\right)\right) \label{trV} \non \\
\tilde V_4&=&g\left(F_4(\hat n)a^2+(a^{\dagger})^2F_4(\hat n) \right)\label{vis}
\eea
and  $F_1$, $F_2$, $F_3$, and $F_4$ are expressed in terms of the
$K$ and $L$ operators as  
\bea
F_1&=&-L(\hat n)L(\hat n+2)\non \\
F_2&=&\frac{1}{2}(1+2\sqrt{2}K(\hat n))\non \\
F_3&=&\frac{1}{2\sqrt{2}}\left(L(\hat n)+L(\hat n+1)(1+2\sqrt{2}K(\hat n))\right)\non \\
F_4&=&\frac{1}{2\sqrt{2}}\left(L(\hat n)-L(\hat n+1)(1+2\sqrt{2}K(\hat n))\right)\non \\
\eea

We can distinguish three types of terms in Eq.(\ref{trV}) based on the physical processes that 
they describe when acting on the $|\theta_n^{\uparrow\downarrow}\ket$ states in Eq. (\ref{jcstates}). 
$\tilde V_1$ describes atomic excitation
or relaxation  though absorption or emission of three photons. $\tilde V_2$
describes the simultaneous excitation of the atom and creation of a photon or 
simultaneous relaxation of 
the atom and absorption of a photon. 
$\tilde{V}_3$ and $\tilde{V}_4$  correspond to creation or annihilation of two
photons with no net change to the excitation state of the atom. 

The question now becomes whether or not keeping only some terms in Eq. (\ref{tildev}) 
leads to a solvable model and 
if so, is there a physical justification for keeping only those terms?
Inspection of Eqs. (\ref{vis}) shows that there are two obvious  cases,
\be
\tilde H_1=\tilde H_{JC} + \tilde V_1, \qquad \tilde H_2=\tilde H_{JC} + \tilde V_2
\ee
The reason for their solvability is the same as for the Jaynes-Cummings model, viz., there exist pairs of states
such that the Hamiltonian can induce transitions only within each pair. 

To determine  if either $\tilde H_1$ or $\tilde H_2$ can be used to approximate $\tilde H$
when describing the system dynamics, we will
use the same approach that justifies the use of the Jaynes-Cummings model
 as an approximation to the total Hamiltonian 
(\ref{Hamrabi}). Thus,
we will write $\tilde H$ in the interaction picture using $\tilde H_{JC}$ as a free Hamiltonian and then
analyze oscillatory behavior for different terms. Within the interaction picture, $\tilde H$ becomes, 
\be
\tilde H_I=e^{i\tilde H_{JC}t}\tilde V e^{-i\tilde H_{JC}t}.
\ee
Note that any operator that depends only on $\hat n$ and 
$\sigma_z$ remains unchanged in the interaction picture. Other operators that appear in 
Eq. (\ref{vis}) have the following interaction picture form
\bea
(\sigma^-(a^{\dagger})^3)_I&=&
\sigma^-(a^{\dagger})^3 e^{iw_1t},   \label{sa3} \\
(\sigma^+a^{\dagger})_I&=&
\sigma^+a^{\dagger} e^{iw_2t}.  \label{sa} \\
(a^{\dagger})^2_I&=&(a^{\dagger})^2 e^{iw_3t},    \label{a2} 
\eea
where
\bea
w_1&=&2\omega -g(\sqrt{\hat n+3}+\sqrt{\hat n+1}), \label{f1} \\
w_2&=&2\omega+g(\sqrt{\hat n+2}+\sqrt{\hat n}), \label{f2}\\
w_3&=&2\omega +g (\sqrt{\hat n}-\sqrt{\hat n+2}) +\frac{g}{2}(\sigma_z+1)\non \\
& &\times(\sqrt{\hat n+3}+\sqrt{\hat n+2}
-\sqrt{\hat n+1}-\sqrt{\hat n}).\label{f3}
\eea
We can see that oscillation frequencies. $\omega_i$ are now operators. 
If we expand the states on which these operators act 
in terms of eigenstates of $\hat n $ and $\sigma_z$ then we can replace 
both $\hat n $ and $\sigma_z$ with
their eigenvalues ($n$ for $\hat n $ and $\pm1$ for $\sigma_z$) 
and $\omega_i$'s 
become $c$-numbers. 
For states 
with moderate occupation number $n$ we can approximate sums of square 
roots in Eqs. (\ref{f1},\ref{f2}) as
\bea
& &\sqrt{n+3}+\sqrt{n+1}\approx2\sqrt{n+1}, \non \\
& &\sqrt{n+2}+\sqrt{n}\approx2\sqrt{n+1}.
\eea 
Differences of square roots in Eq. (\ref{f3}) are of order $1/\sqrt{n}$
and the terms involving these differences can be omitted if
$g/\sqrt{n}\ll\omega$.
This gives the following approximation for the effective frequencies
\bea
w_1&\approx&
2(\omega-g\sqrt{n+1})  \\ \label{sa3ap} 
w_2&\approx&
2(\omega+g\sqrt{n+1})  \\ \label{saap}
w_3&\approx&2\omega  \label{a2ap} 
\eea
Comparing these effective frequencies, we can see the for $g>0$  
we have $2|(\omega-g\sqrt{n+1})|<2\omega<2(\omega+g\sqrt{n+1})$
if $g\sqrt{n+1}<2\omega$. In this case operators appearing in Eq. (\ref{sa3})
will have the slowest oscillating frequency.  
Similarly, if $g<0$,
operators in Eq. (\ref{sa}) will have slower oscillating frequency then operators 
(\ref{sa3}) and (\ref{a2}) if $-g\sqrt{n+1}<2\omega$ is satisfied.
Thus, we may expect $\tilde H_1$ to give a reasonable description of the system dynamics
for positive $g$ and $\tilde H_2$ for negative $g$ for specified ranges of parameters. 
Even though these approximations may be 
unsatisfactory in other regimes, we anticipate that  some 
of the complex system dynamics 
that is present in the Rabi Hamiltonian will be manifest  in 
our approximating Hamiltonians.
  
We now recall that the sign of $g$ in 
Eq. (\ref{Hamrabi}) can always be chosen as either negative or positive 
without the loss of generality. This is because there are two unitary
transformation whose action on Hamiltonian (\ref{Hamrabi})
is equivalent to changing the sign of $g$. One is the space inversion transformation
which changes the sign of $a$ and $a^{\dagger}$ and leaves $a^{\dagger}a$ invariant.
(This transformation is generated
by  $\exp{(i \pi a^{\dagger}a)}$). Another is the transformation generated by 
$\exp{(i \frac{\pi}{2}(\sigma_z+1) )}$ that changes the sign of $\sigma^+$ and $\sigma^-$
but leaves $\sigma_z$ invariant.
Thus we can approximate Hamiltonian $\tilde H$ by either $\tilde H_1$ or
$\tilde H_2$ depending on our choice of sign for $g$.

\section{Eigenstates and eigenvalues of the approximating Hamiltonians} \label{Solution}

\begin{figure}
 \includegraphics[width=\columnwidth]{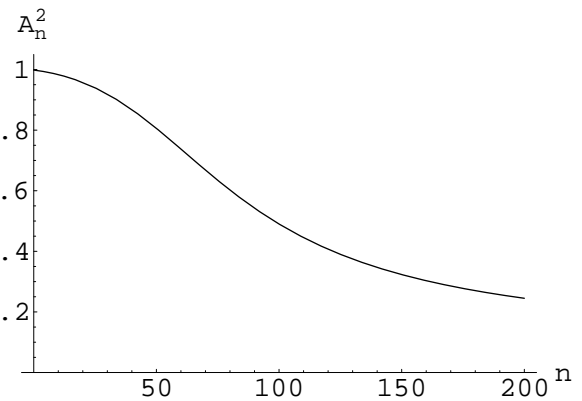} 
 \caption{\label{A2vsn}$A^2(n)$ as a continuous function of $n$ for
 weak coupling ($\omega=1, g=0.1$).}
 \end{figure}

First, we will consider Hamiltonian $\tilde H_1$. Its eigenstates and eigenvalues can be
found along the same lines as for the Jaynes-Cummings model, i.e. by diagonalizing suitable
 two by two matrices. 
The eigenstates have the form
\bea
|\tilde\phi^-_n\ket&=&A_n|\uparrow\ket|n\ket +B_n |\downarrow\ket|n+3\ket,\,\,\,\, {\rm and}\non \\
|\tilde\phi^+ _n\ket&=&B_n|\uparrow\ket|n\ket -A_n |\downarrow\ket|n+3\ket.\label{states1}
\eea
Here $n\geq0$ and 
\be
A_n=\frac{1}{\sqrt{1+\alpha_n^2}}, \qquad B_n=\frac{\alpha_n}{\sqrt{1+\alpha_n^2}} \label{an},
\ee
where
\bea
\alpha_n&=&\frac{\mu_n}{\Delta_n+\eta_n}, \non \\
\mu_n&=&g\sqrt{n+2}, \non \\
\eta_n&=&2\omega-g(\sqrt{n+1}+\sqrt{n+3}), \non \\
\Delta_n&=&\sqrt{\mu_n^2+\eta_n^2}. \label{coefficients1}
\eea
Eigenvalues corresponding to eigenstates (\ref{states1}) are
\be
\kappa^{\pm}_n=\frac{1}{2}((2n+3)\omega+g(\sqrt{n+1}-\sqrt{n+3})\pm \Delta_n). \label{values1}
\ee
Fig.\ref{A2vsn} gives  $A^2(n)$ plotted as a continuous function of $n$ in the case
weak coupling. It can be seen that for low-lying values of $n$, 
$A^2(n)$ is close to one indicating that in this region eigenstates  (\ref{states1}) 
are similar to the Jaynes-Cummings eigenstates. 

In addition to eigenstates (\ref{states1}), there are three special eigenstates 
of $\tilde H_1$ with eigenvalues
\be
k_0=-\frac{\om}{2}, \qquad k_1=\frac{\om}{2}-g, \qquad k_2=\frac{3\om}{2}-\sqrt{2}g \label{spvalues1}
\ee
All three are also eigenstates of $\tilde H_{JC}$ and are given by
\be
|\tilde\chi_0\ket=|\downarrow\ket|0\ket, \qquad
|\tilde\chi_1\ket=|\downarrow\ket|1\ket ,\qquad
|\tilde\chi_2\ket=|\downarrow\ket|2\ket . \label{special1}
\ee

Moving on, we now consider eigenstates and eigenvalues of $\tilde H_2$.
Its eigenstates have the form
\bea
{|\tilde\psi^-_n\ket}&=&C_n|\downarrow\ket|n+1\ket +D_n |\uparrow\ket|n+2\ket,\non \\
{|\tilde\psi^+_n\ket}&=&D_n|\downarrow\ket|n+1\ket -C_n |\uparrow\ket|n+2\ket \label{states2}
\eea
Here $n\geq0$.
We denote eigenvalues corresponding to states $|\phi_n^{\pm}\ket$ by $\la_n^{\pm}$. 
Remarkably, the following relationship holds between coefficients $C_n$ and $A_n$ as well as
$D_n$ and $B_n$ viewed as functions of the coupling parameter $g$,
\be
C_n(g)=A_n(-g), \qquad D_n(g)=B_n(-g) \label{relation1}
\ee
Similarly, we have for eigenvalues 
\be
\lambda_n^{\pm}(g)=\kappa_n^{\pm}(-g). \label{relation2}
\ee
The validity of relations (\ref{relation1}) and (\ref{relation2}) can be easily checked if 
one writes for $\tilde H_1$ and $\tilde H_2$ explicit $2\times2$ matrices whose diagonalization 
gives coefficents in Eq.(\ref{states1}) and Eq.(\ref{states2}) and eigenvalues for $\tilde H_1$ and $\tilde H_2$. 
For a given $n$, these matrices only differ by the sign in front of $g$.

As in the case of Hamiltonian $\tilde H_1$, there are three additional special 
eigenstates of $\tilde H_2$. Two of them have the same
general form as eigenstates (\ref{states2}) but they do not have
any simple relation (such as Eqs. (\ref{relation1},\ref{relation2}))
to the special states of $\tilde H_1$. These two eigenstates are 
\be
|\tilde\xi ^-\ket=c|\downarrow\ket|0\ket +d |\uparrow\ket|1\ket,\qquad
|\tilde\xi^+\ket=d|\downarrow\ket|0\ket -c |\uparrow\ket|1\ket \label{specialstates2}
\ee
Here, $c$ and  $d$ are given by
\be
c=\frac{1}{\sqrt{1+\gamma^2}},\qquad d=\frac{\gamma}{\sqrt{1+\gamma^2}},
\ee
where
\bea
\gamma&=&-\frac{g}{g+\sqrt{2}(\omega+\epsilon)},\non \\
\epsilon&=&\sqrt{g^2+\sqrt{2}g\om +\om^2}. 
\eea
The corresponding eigenvalues are 
\be
l^{\pm}=\frac{1}{2}(\om+\sqrt{2}g\pm2\epsilon).
\ee
The third special state of $\tilde H_2$ is also an eigenstate of $\tilde H_{JC}$. It is given by
\be
|\tilde\xi_0\ket=|\uparrow\ket|0\ket
\ee
with eigenvalue 
\be
l_0=\frac{\om}{2}+g
\ee

With the knowledge of eigenstates and eigenvalues of approximating Hamiltonians
$\tilde H_1$ and $\tilde H_2$ we can calculate the time evolution of any
observable. 
However, since observables of interest and initial states are given in the original 
untransformed picture, it is convenient to remain in this picture, in which case the time 
evolution is determined by 
\be
H_1=U^{-1}\tilde H_1U, \qquad H_2=U^{-1}\tilde H_2 U \label{transH}
\ee
Explicit operator forms for $H_1$ and $H_2$ are given in the Appendix.
Eigenstates of these operators are obtained by acting with $U^{-1}$ on states
 (\ref{states1},\ref{special1}) in case
of $H_1$ or  states (\ref{states2},\ref{specialstates2}) in case of $H_2$. 
Eigenstates of  $H_1$ are 
\bea
{|\phi^-_n\ket}&=&\frac{1}{\sqrt{2}}|\downarrow\ket
\left(A_n|n+1\ket +B_n|n+3\ket\right) \non \\
& &+\frac{1}{\sqrt{2}}|\uparrow\ket
\left(-B_n|n+2\ket +A_n|n\ket\right), \non \\
{|\phi^+_n\ket}&=&\frac{1}{\sqrt{2}}|\downarrow\ket
\left(B_n|n+1\ket -A_n|n+3\ket\right) \non \\
& &+\frac{1}{\sqrt{2}}|\uparrow\ket
\left(A_n|n+2\ket +B_n|n\ket\right). \label{trstates1}
\eea
The special states of  $H_1$ are given by 
\bea
|\chi_0\ket&=&|\downarrow\ket|0\ket,\non \\
|\chi_1\ket&=&\frac{1}{\sqrt{2}}\left(|\downarrow\ket|1\ket-|\uparrow\ket|0\ket \right),\non \\
|\chi_2\ket&=&\frac{1}{\sqrt{2}}\left(|\downarrow\ket|2\ket-|\uparrow\ket|1\ket \right).\label{trspecial1}
\eea
For eigenstates of  $H_2$ we have
\bea
{|\psi^-_n\ket}&=&\frac{1}{\sqrt{2}}|\downarrow\ket
\left(C_n|n+1\ket +D_n|n+3\ket\right) \non \\
& &+\frac{1}{\sqrt{2}}|\uparrow\ket
\left(D_n|n+2\ket -C_n|n\ket\right), \non \\
{|\psi^+ _n\ket}&=&\frac{1}{\sqrt{2}}|\downarrow\ket
\left(D_n|n+1\ket -C_n|n+3\ket\right) \non \\
& &-\frac{1}{\sqrt{2}}|\uparrow\ket
\left(C_n|n+2\ket +D_n|n\ket\right).\label{trstates2}
\eea
The special eigenstates are
\bea
|\xi^-\ket&=&|\downarrow\ket\left(c|0\ket+\frac{d}{\sqrt{2}}|2\ket\right)
+\frac{d}{\sqrt{2}}|\uparrow\ket|1\ket,\non \\
|\xi^+\ket&=&|\downarrow\ket\left(d|0\ket-\frac{c}{\sqrt{2}}|2\ket\right)
-\frac{c}{\sqrt{2}}|\uparrow\ket|1\ket, \non \\
|\xi_0\ket&=&\frac{1}{\sqrt{2}}|\downarrow\ket|1\ket
+\frac{1}{\sqrt{2}}|\uparrow\ket|0\ket.
\eea

We can see that each of the states (\ref{trstates1}) and (\ref{trstates2}) is a superposition
of {\it{four}} eigenstates of operators $\sigma_z$ and $\hat n$.
In contrast, we may recall that eigenstates of $H_{JC}$ are superpositions of only 
{\it{two}} such states. 

\section{Dynamics of the atomic survival probability} \label{Survival}

We will now consider time evolution of the probability $P(t)$ for the atom 
to be exited if the initial state
of the system given by  $|\uparrow\ket |f\ket=|\uparrow,f\ket$ where 
$|f\ket$ is an arbitrary state of the field. $P(t)$ is expressed in terms 
of $\bra\sigma_z(t)\ket$ as 
\be
P(t)=\frac{1}{2}(1+\bra\sigma_z(t)\ket).
\ee
Let us consider the case of Hamiltonian $H_1$.
Straightforward calculations using a complete set of eigenstates 
(\ref{trstates1}, \ref{trspecial1}) give for $P(t)$ 
\bea
P(t)&=&\frac{1}{2}+   \Re e \Bigg[\sum_{n=0}^{\infty}\left(A_nA_{n+2}{F^-_{n+2}}^*{F^+_n}
e^{i(\kappa^-_{n+2}-\kappa^+_{n})t}\right. \non \\ 
& & \left.-B_nA_{n+2}{F^-_{n+2}}^*{F^-_n}
e^{i(\kappa^-_{n+2}-\kappa^-_{n})t}  \right. \non \\
& &\left.\left. +A_nB_{n+2}{F^+_{n+2}}^*{F^+_n}
e^{i(\kappa^+_{n+2}-\kappa^+_{n})t} \right.\right. \non \\
& &-\left.\left. B_nB_{n+2}{F^+_{n+2}}^*F_n^-
e^{i(\kappa^+_{n+2}-\kappa^-_n)t}\right) \right.\non \\
& &\left.+\frac{1}{\sqrt{2}}\left(A_0\bra 0|f\ket{F^-_0}^*e^{i(\kappa^-_0-k_1)t} \right.\right.\non \\
& &\left.\left.+B_0\bra 0|f\ket{F^+_0}^*e^{i(\kappa^+_0-k_1)t} \right.\right. \non \\
& &\left.\left.+A_1\bra 1|f\ket{F^-_1}^*e^{i(\kappa^-_1-k_2)t} \right.\right. \non \\
& &\left.+B_1\bra 1|f\ket{F^+_1}^*e^{i(\kappa^+_1-k_2)t}\right)\Bigg]. \label{population}
\eea
Here 
\bea
F^{-}_n&=&\bra\phi^{-}_n|\uparrow,f\ket
=\frac{1}{\sqrt{2}}(A_n\bra n|f\ket-B_n\bra n+2|f\ket), \\  \label{overlap1}
F^{+}_n&=&\bra\phi^{+}_n|\uparrow,f\ket
=\frac{1}{\sqrt{2}}(A_n\bra n+2|f\ket+B_n\bra n|f\ket).  \\ \label{overlap2}
\eea

In order to qualitatively understand time dependence of $P(t)$ 
let us classify contributions from various terms in Eq. (\ref{population}). All the terms
in brackets have the form of time dependent exponentials preceded 
by a factor. Absolute values
of these factors depend on the initial state of the field.
The term in the second parentheses is due to the overlap of the initial state $|\uparrow,f\ket$
with the special states (\ref{trspecial1}). Its contribution is negligible for initial states
with small $\bra 0|f\ket$, $\bra 1|f\ket$, $\bra 2|f\ket$, and $\bra 3|f\ket$ components. 
Oscillating exponentials that appear in the first parentheses can be divided into two
groups - those involving differences of eigenvalues with the same superscripts 
and those involving differences of eigenvalues with the different superscripts.

Let us consider the $n$ dependence of $\kappa^+_{n+2}-\kappa^+_{n}$. 
We will again assume that
states $|f\ket$ have not too small average $\hat n$. Using
explicit form of eigenvalues given by Eq. (\ref{values1}) it can be shown that 
\be
\kappa^+_{n+2}-\kappa^+_{n}=2\omega+{\cal O}\left(\frac{g}{\sqrt{n+1}}\right)
\ee
Hence, this difference can be approximated by
$2\omega$ when $g/\sqrt{n+1}\ll \omega$ which holds for many 
couplings and initial states of interest. 
A similar result holds for
$\kappa^-_{n+2}-\kappa^-_{n}$. Thus, the second and third terms in the 
first parentheses in  Eq. (\ref{population}) give primarily oscillating 
contributions with frequency of about $2\omega$.

Eigenvalue differences $\kappa^-_{n+2}-\kappa^+_{n}$ and 
$\kappa^+_{n+2}-\kappa^-_{n}$
have more complicated $n$ dependences.  
It can be shown that for states with typical $n\gg1$ but such that
$g\sqrt{n+1}\ll\omega$ 
\bea
\kappa^-_{n+2}-\kappa^+_{n}&\approx& 2g\sqrt{n+1},\label{difference}\\
\kappa^+_{n+2}-\kappa^-_{n}&\approx&4\omega-2g\sqrt{n+1}.
\eea
These expressions allow to make connection with the
standard Jaynes-Cummings model. 

We showed earlier that for weak coupling  
coefficients $A_n$ are close to one and, therefore, $B_n$ are close 
to zero for moderate values of  $n$ (Fig.\ref{A2vsn}).
Thus, the first term in the first parentheses in Eq. (\ref{population})
is dominant for weak coupling for states with moderate average $\hat{n}$ values.
Replacing  $A_n\to1$  and  $B_n\to0$ in Eqs.(\ref{overlap1},\ref{overlap2}), 
approximating $\bra n+2|f\ket$ with $\bra n|f\ket$, using  
Eq. (\ref{difference}), and neglecting the terms in the second parentheses in 
Eq. (\ref{population}) we obtain the survival probability for
the resonant Jaynes-Cummings model \cite{Shore}
\be
P(t)\approx\frac{1}{2}\left(1+
\sum_0^\infty|\bra n|f\ket|^2\cos(2g\sqrt{n+1}\,t)\right).
\ee  

Using Eq. (\ref{population}), we can compare the survival probability $P_{AHM}(t)$
from our approximating Hamiltonian model to $P_{JCM}(t)$ from the Jaynes-Cummings model
 as well
as the exact survival $P_{RH}(t)$ for the Rabi Hamiltonian 
(obtained by exact numerical integration of the
corresponding Schr\"{o}dinger equation). We treat both the Jaynes-Cummings model 
and the approximating Hamiltonian model as 
approximations of the Rabi Hamiltonian.

\begin{figure}
 \includegraphics[width=\columnwidth]{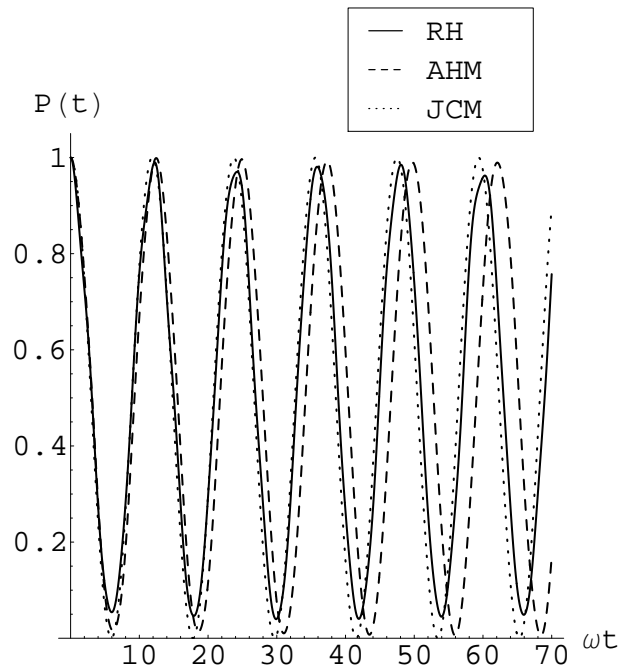} 
 \caption{\label{Num6}Excited state survival probability 
 for the Rabi Hamiltonian (RH), the approximating Hamiltonian model (AHM), 
 and the Jaynes-Cummings model (JCM) ($\omega=1, g=0.1$) for 
 the number initial state of 
 the field with $n=6$.}
 \end{figure}

Fig. \ref{Num6} shows survival probabilities $P(t)$ for the three models
for the weak coupling case and when the initial state of the field is the number state
with $n=6$. All models show qualitatively similar
behavior of the Rabi type oscillations. However, $P_{RH}(t)$ never completely collapses
 (Fig. \ref{Num6}). This effect, although not so pronounced, is visible in the
 approximating Hamiltonian model as well. It can also be seen that the Rabi frequency
  for the Jaynes-Cummings model
 is very slightly larger
then the oscillation frequency for the Rabi Hamiltonian whereas 
for the approximating Hamiltonian model
 it is slightly 
smaller.

\begin{figure}
 \includegraphics[width=\columnwidth]{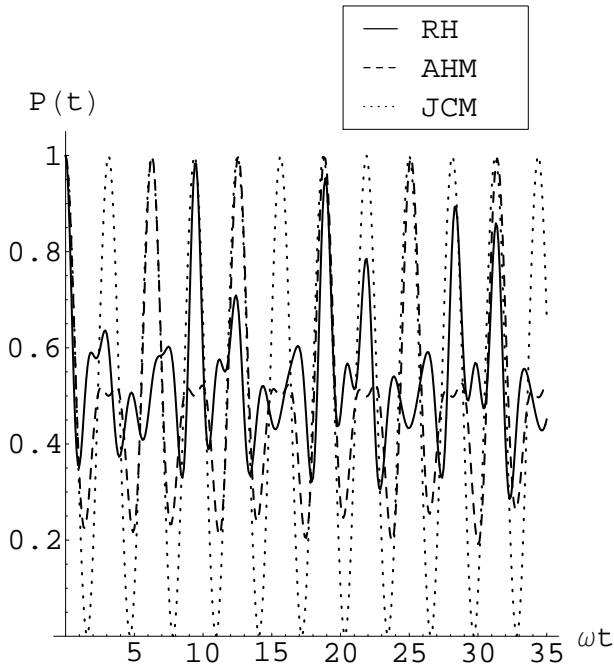} 
 \caption{\label{Num100}Excited state survival probability 
 for the Rabi Hamiltonian, the approximating Hamiltonian model, 
 and the Jaynes-Cummings model ($\omega=1, g=0.1$) for 
 the number initial state of 
 the field with $n=100$.}
 \end{figure}

Fig. \ref{Num100}  gives survival 
probabilities for the  weak coupling case with the initial state of the field
taken as the number state
with $n=100$. We can see that in this case of the strong
field both the Jaynes-Cummings model and the approximating Hamiltonian model
 deviate from the Rabi Hamiltonian. However, qualitatively,
the approximating Hamiltonian model gives a better description. As the Rabi Hamiltonian, 
the approximating Hamiltonian model shows absence of complete collapse
and strong deviation from simple oscillatory behavior.

\begin{figure}
 \includegraphics[width=\columnwidth]{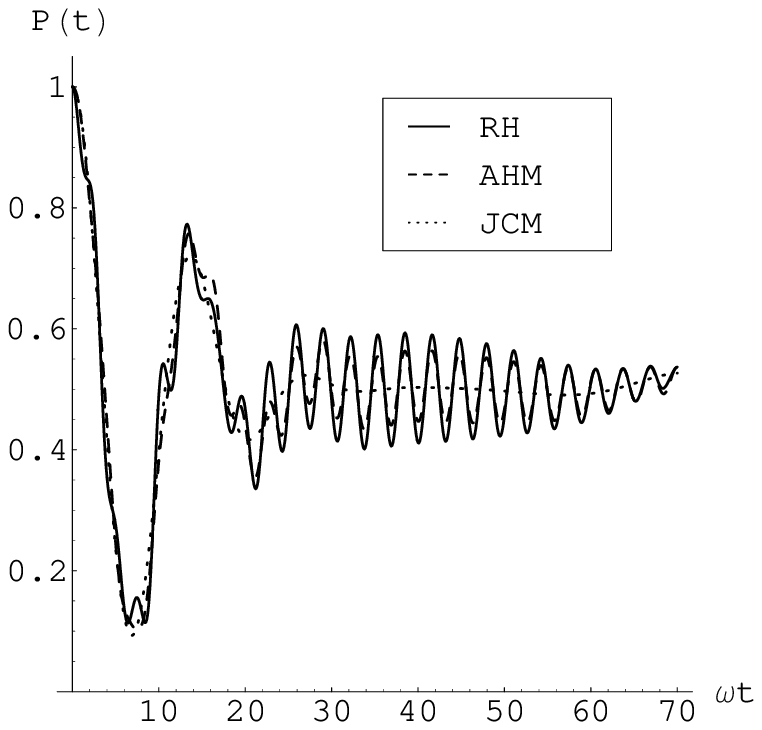} 
 \caption{\label{Coh4}Excited state survival probability 
 for the Rabi Hamiltonian, the approximating Hamiltonian model, and the 
 Jaynes-Cummings model ($\omega=1, g=0.1$) for the coherent initial state of 
 the field with $\alpha=2, (\bra n\ket=4)$.}
 \end{figure}

Taking the initial state of the field as a coherent leads to the results shown on 
Fig. \ref{Coh4} in the 
case of the weak coupling and weak field ($\bra n\ket=4$). The approximating Hamiltonian model 
approximates 
the Rabi Hamiltonian better then the Jaynes-Cummings model because it account for 
fast oscillations in $P(t)$ with
the frequency of about $2\omega$. However, the intensity of these oscillations
is weaker compared to the Rabi Hamiltonian. 

\begin{figure}
 \includegraphics[width=\columnwidth]{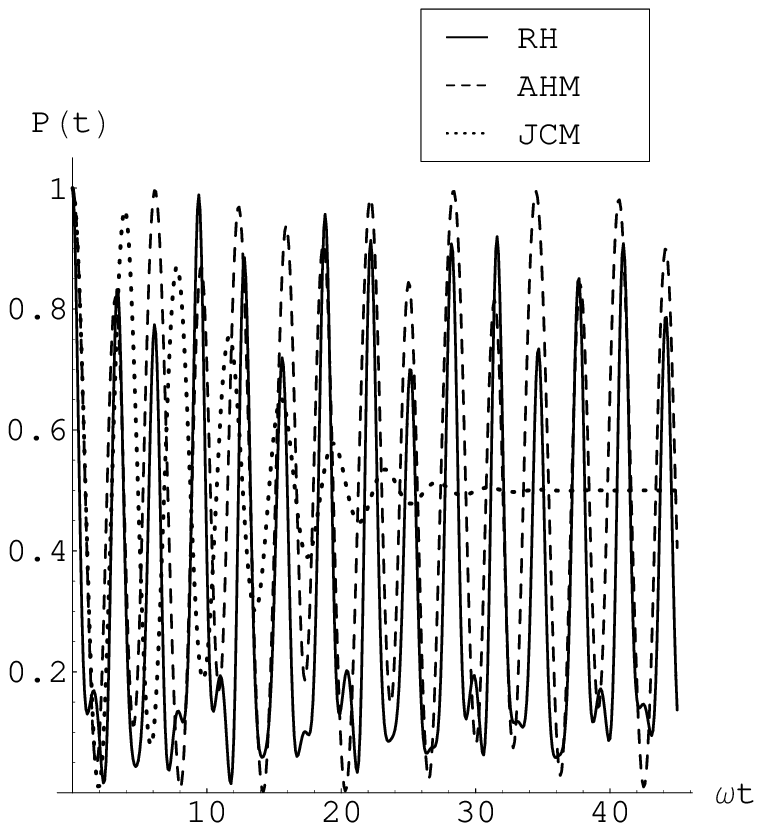} 
 \caption{\label{Coh64}Excited state survival probability 
 for the Rabi Hamiltonian, the approximating Hamiltonian model, and the Jaynes-Cummings 
 model ($\omega=1, g=0.1$) for the coherent initial state of 
 the field with $\alpha=8, (\bra n\ket=64)$.}
 \end{figure}

In the case of the strong coherent initial field 
(Fig. \ref{Coh64}) 
The approximating Hamiltonian model again gives a better approximation to 
the Rabi Hamiltonian then the Jaynes-Cummings model.
Both the Rabi Hamiltonian and the approximating Hamiltonian model show 
almost periodic revivals of
$P(t)$ with no apparent weakening. In contrast to the Jaynes-Cummings model, both models
do not have a region of nearly constant $P(t)$ before the 
onset of the second group of collapses and revivals that is present in 
the Jaynes-Cummings model.

It is easy construct explicit expression for $P(t)$ in the case of 
Hamiltonian $H_2$ using its eigenstates and eigenvalues. 
One can expect that due to the symmetry properties given by Eqs. 
(\ref{relation1},\ref{relation2}) the following relationship will hold
between $P(t)$ for $H_1$ and $P(t)$ for $H_2$ viewed as functions of $g$
\be
P_{H_1}(g,t)\approx P_{H_2}(-g, t).
\ee
In general, the equality is not exact due contributions form the special states
for Hamiltonians $H_1$ and $H_2$ for which there are no symmetry relations. 
We will not pursue investigation of $P_{H_2}(t)$ here since for $g<0$ it gives 
results that are qualitatively similar to $P_{H_1}$ for $g>0$. 

\section{Conclusions}
In this paper we present an approach that allows one to add extra
terms to the Jaynes-Cummings model of an atom in an external field. These
additional terms add complex oscillatory terms to the survival probability which 
become increasingly important as the field intensity is increased. By retaining 
select portions of the full counter-rotating term from the original 
Rabi Hamiltonian we obtain an analytically solvable model that compares favorably 
with the numerically exact survival probabilities from the Rabi Hamiltonian
over a wide range of parameters even for relatively strong field strengths.

\appendix* 
\section{}
Explicit forms of Hamiltonians $H_1$ and $H_2$ are obtained by using
Eqs. (\ref{unitary},\ref{transH})
\bea
H_1&=&H_{JC}+\bigg(-\frac{g}{2}\sigma^+L(\hat{n})L(\hat{n}+2)a^3
+\frac{g}{4}\sigma^-\big(1-\delta(\hat{n})\big)a\non \\
& &+\frac{g}{4\sqrt{2}}(1+\sigma_z)L(\hat{n})a^2  \non \\
& &-\frac{g}{4\sqrt{2}}(1-\sigma_z)\big(1-\delta(\hat{n})\big)L(\hat{n}+1)a^2+h.c.\bigg)  \\
H_2&=&H_{JC}+\bigg(-\frac{g}{2}\sigma^+L(\hat{n})L(\hat{n}+2)a^3
+\frac{g}{4}\sigma^-\big(1+\delta(\hat{n})\big)a\non \\
& &-\frac{g}{4\sqrt{2}}(1+\sigma_z)L(\hat{n})a^2  \non \\
& &+\frac{g}{4\sqrt{2}}(1-\sigma_z)\big(1+\delta(\hat{n})\big)L(\hat{n}+1)a^2+h.c.\bigg)  
\eea
where operator $L(\hat{n})$ is defined in Eq. (\ref{kl}) and $h. c$ stands for 
Hermitian conjugate. We can see that in Hamiltonians $H_1$ 
and $H_2$, the counter-rotating terms are replaced by a number of terms
involving various intensity dependent multi-photon transitions. 

\begin{acknowledgments}
This work was funded in part through grants from the National Science
Foundation and the Robert A. Welch foundation.
\end{acknowledgments}

\end{document}